\documentstyle[11pt]{article}

\begin{document}

\title{\vspace*{-1cm}
{\bf Improvement of measurement accuracy in SU(1,1) interferometers}}
\author{\Large{C. Brif \thanks{E-mail: costya@physics.technion.ac.il}
\ and \ Y. Ben-Aryeh} \vspace*{0.2cm} \\
{\em Department of Physics, Technion -- Israel Institute of 
Technology, Haifa 32000, Israel}}
\date{published in {\em Quantum \& Semiclassical Optics} 
{\bf 8}, 1 (1996) }
   \maketitle

	\begin{abstract}
\noindent
We consider an SU(1,1) interferometer employing four-wave mixers 
that is fed with two-mode states which are both coherent and 
intelligent states of the SU(1,1) Lie group. It is shown that the
phase sensitivity of the interferometer can be essentially improved
by using input states with a large photon-number difference between
the modes.
	\end{abstract}
\vspace*{0.4cm}

\noindent
The improvement of measurement accuracy in interferometers is of
significant importance in modern experimental physics. Much work has 
been done on the reduction of the quantum noise in interferometers 
by using input light fields prepared in nonclassical photon states.
It was pointed out by Caves \cite{Caves} and Bondurant and Shapiro 
\cite{BoSh} that the quantum fluctuations can be diminished by 
feeding squeezed states of light into the interferometer. The 
interferometers considered in \cite{Caves,BoSh} employ passive 
lossless devices, such as beam splitters. Yurke, McCall and Klauder 
\cite{YMK} showed that such interferometers can be characterized by 
the SU(2) group. They also introduced a class of interferometers 
which employ active lossless devices, such as four-wave mixers, and 
are characterized by the SU(1,1) group. It was shown \cite{YMK} that 
the use of squeezed light in SU(2) interferometers can yield a phase 
sensitivity $\Delta\phi \sim 1/N$ (where $N$ is the total number of 
photons passing through the interferometer), while SU(1,1) 
interferometers can achieve a phase sensitivity of $1/N$ with only 
vacuum fluctuations entering the input ports.

In the present work we study the possibility to improve further the
accuracy of SU(1,1) interferometers by using specially prepared
states (other than vacuum). We apply the idea of Hillery and Mlodinow
\cite{HM} who proposed to use intelligent states (IS) \cite{Arag}
for improving the phase sensitivity of interferometers. They analysed
\cite{HM} the case of SU(2) IS. Since we discuss here interferometers
characterized by SU(1,1), it is natural to use IS of this group
\cite{WE,BBA,GeGr}.
There is a problem of generating IS since, in general, they are
constructed by nonunitary operators \cite{Arag,WE}. However, there 
are some IS which simultaneously are generalized coherent states (CS)
\cite{Per,Gil} of the corresponding Lie group, i.e., an intersection 
occurs between these two types of states \cite{WE}. This intersection
is of special importance in physics because IS that also are CS can
be created by Hamiltonians for which a given Lie group is the
dynamical symmetry group. Recently we developed \cite{BBA} a general
group-theoretical approach to SU(1,1) IS by representing them in the 
corresponding coherent-state basis. This approach yields the most 
full characterization of the coherent-intelligent intersection. The 
above results will be used in the present work for analysing SU(1,1) 
interferometers fed with states which are both IS and CS of the 
SU(1,1) Lie group.

An SU(1,1) interferometer is described schematically in figure 1. 
Two light beams represented by mode annihilation operators $a_{1}$ 
and $a_{2}$ enter into the input ports of the first four-wave mixer 
FWM1. After leaving FWM1, beams accumulate phase shifts $\phi_{1}$
and $\phi_{2}$, respectively, and then they enter the second
four-wave mixer FWM2. The photons leaving the interferometer are
counted by detectors D1 and D2. 

For the analysis of such an interferometer it is convenient to 
consider the Hermitian operators
\begin{equation} 
K_{1} =\frac{1}{2}(a_{1}^{\dagger}a_{2}^{\dagger} + a_{1}a_{2}) , 
\;\;\;\;\;\;\;
K_{2} =\frac{1}{2i}(a_{1}^{\dagger}a_{2}^{\dagger} - a_{1}a_{2}) , 
\;\;\;\;\;\;\;
K_{3} =\frac{1}{2}(a_{1}^{\dagger}a_{1} + a_{2}a_{2}^{\dagger}) .
\label{1} \end{equation}
These operators form the two-mode boson realization of the SU(1,1)
Lie algebra:
\begin{equation} 
[ K_{1}, K_{2} ] = -iK_{3} ,   \;\;\;\;\;\;\; 
[ K_{2}, K_{3} ] = iK_{1} ,    \;\;\;\;\;\;\;
[ K_{3}, K_{1} ] = iK_{2} .
\label{2}   \end{equation}
It is also useful to introduce raising and lowering operators
\begin{equation} 
K_{+} = K_{1} + iK_{2} = a_{1}^{\dagger}a_{2}^{\dagger} , 
\;\;\;\;\;\;\;
K_{-} = K_{1} - iK_{2} = a_{1}a_{2} .
\label{3}   \end{equation}
The Casimir operator
\begin{equation}
K^{2} = K_{3}^{2} - K_{1}^{2} - K_{2}^{2}
\label{4}  \end{equation}
for any unitary irreducible representation is the identity operator 
$I$ times a number:
\begin{equation}
K^{2} = k(k-1)I .
\label{5}  \end{equation}
Thus a representation of SU(1,1) is determined by a single number
$k$ that is called Bargmann index. For the discrete-series
representations \cite{SU11} the Bargmann index acquires discrete 
values $k=\frac{1}{2},1,\frac{3}{2},2,\ldots$.
By using the operators (\ref{1}), one gets
\begin{equation}
K^{2} = \frac{1}{4}(a_{1}^{\dagger}a_{1} - a_{2}^{\dagger}a_{2})^{2}
- \frac{1}{4} . \label{6}  \end{equation}
The photon-number difference between the modes
$n_{0} = \langle a_{1}^{\dagger}a_{1} - a_{2}^{\dagger}a_{2}
\rangle$ is a constant (chosen to be positive) for each irreducible 
representation and it is related to the Bargmann index via
$k=\frac{1}{2}(n_{0}+1)$.
The corresponding state space is spanned by the complete orthonormal
basis $|k,n\rangle$ $(n=0,1,2,\ldots)$ that can be expressed
in terms of Fock states of two modes:
\begin{equation}
|k,n\rangle = |n+n_{0}\rangle_{1} |n\rangle_{2} .
\label{9}  \end{equation}

The actions of the interferometer elements on the vector
${\bf K}=(K_{1},K_{2},K_{3})$ can be represented as Lorentz boosts
and rotations in the (2+1)-dimensional space-time \cite{YMK}. 
FWM1 acts on ${\bf K}$ as a Lorentz boost with the transformation
matrix
\begin{equation} 
L(-\beta) =\left( \begin{array}{ccc} 
1 & 0 & 0 \\ 
0 & \cosh\beta & - \sinh\beta \\
0 & - \sinh\beta & \cosh\beta   \end{array}  \right) . 
\label{10}  \end{equation}
The transformation matrix of FWM2 is $L(\beta)$, i.e., two four-wave
mixers perform boosts in opposite directions. Phase shifters rotate
${\bf K}$ about the 3rd axis by an angle $\phi=-(\phi_{1}+
\phi_{2})$. The transformation matrix of this rotation is
\begin{equation} 
R(\phi) =\left( \begin{array}{ccc}
\cos\phi & - \sin\phi & 0 \\
\sin\phi & \cos\phi & 0 \\
0 & 0 & 1          \end{array}  \right) . 
\label{11}  \end{equation}
The overall transformation performed on ${\bf K}$ is
\begin{equation}
{\bf K}_{{\rm out}} = L(\beta)R(\phi)L(-\beta) {\bf K} .
\label{12}  \end{equation}

The information on $\phi$ is inferred from the photon statistics of 
the output beams. One should measure the total number of photons in 
the two output modes, $N_{{\rm out}}$, or, equivalently, the operator 
$K_{3\,{\rm out}} = \frac{1}{2}(N_{{\rm out}}+ 1)$.
The mean-square fluctuation in $\phi$ due to the photon statistics
is given by \cite{YMK}
\begin{equation}
(\Delta\phi)^{2} = \frac{ (\Delta K_{3\,{\rm out}})^{2} }{ \left|
\partial\langle K_{3\,{\rm out}}\rangle /\partial\phi \right|^{2}} .
\label{13}  \end{equation}
From Eq.\ (\ref{12}), we find
\begin{equation}
K_{3\,{\rm out}} = (\sinh\beta \sin\phi) K_{1} 
+ \sinh\beta \cosh\beta (\cos\phi - 1) K_{2} 
+ (\cosh^{2}\!\beta - \sinh^{2}\!\beta \cos\phi) K_{3} .
\label{14}  \end{equation}
If only vacuum fluctuations enter the input ports, then Eq.\ 
(\ref{13}) with $K_{3\,{\rm out}}$ of form (\ref{14}) reduces to the
known result \cite{YMK}
\begin{equation}
(\Delta\phi)^{2} = \frac{ \sin^{2}\!\phi + \cosh^{2}\!\beta (1 - 
\cos\phi)^{2} }{ \sin^{2}\!\phi \sinh^{2}\!\beta } .
\label{15}  \end{equation}
For $\phi=0$ the $(\Delta\phi)^{2}$ is minimized, $(\Delta\phi)^{2}
= 1/\sinh^{2}\!\beta$.

We would like to investigate a more general case when the 
interferometer is fed with an SU(1,1) intelligent state. The 
motivation 
for using IS is as follows. By putting $\phi=0$, we can simplify Eq.\
(\ref{13}) with $K_{3\,{\rm out}}$ given by (\ref{14}) to the form
\begin{equation}
(\Delta\phi)^{2} = \frac{ (\Delta K_{3})^{2} }{ \sinh^{2}\!\beta
|\langle K_{1} \rangle|^{2} } .
\label{16}  \end{equation}
The commutation relation $[K_{2},K_{3}] = iK_{1}$ implies the
uncertainty relation
\begin{equation}
(\Delta K_{2})^{2} (\Delta K_{3})^{2} \geq \frac{1}{4} 
|\langle K_{1} \rangle|^{2} .    \label{17}  \end{equation}
Therefore,
\begin{equation}
(\Delta\phi)^{2}\geq\frac{1}{4\sinh^{2}\!\beta(\Delta K_{2})^{2}} .
\label{18}  \end{equation}
For IS an equality is achieved in the uncertainty relation.
Such $K_{2}$-$K_{3}$ IS with large values
of $\Delta K_{2}$ would allow us to measure small changes in
$\phi$. For these states Eq.\ (\ref{18}) reads
\begin{equation}
(\Delta\phi)^{2} =\frac{1}{ 4\sinh^{2}\!\beta (\Delta K_{2})^{2} } .
\label{19}  \end{equation}
The $K_{2}$-$K_{3}$ IS $|\lambda\rangle_{23}$ are determined from 
the eigenvalue equation 
\begin{equation}
(K_{2} + i\gamma K_{3}) |\lambda\rangle_{23} = \lambda 
|\lambda\rangle_{23} ,     \label{20}  \end{equation}
where $\lambda$ is a complex eigenvalue and $\gamma$ is a real 
parameter given by
\begin{equation} 
|\gamma| = \Delta K_{2}/ \Delta K_{3} . 
\label{21}  \end{equation}
For $|\gamma| > 1$ IS are squeezed in $K_{3}$ and for $|\gamma| < 1$ 
IS are squeezed in $K_{2}$. 

In order to be able to create IS, we must choose states which lie in 
the intersection of the SU(1,1) intelligent and coherent states.
The generalized SU(1,1) CS were introduced by Perelomov \cite{Per}:
\begin{eqnarray}
|k,\zeta\rangle & = & \exp(\xi K_{+} - \xi^{\ast} K_{-}) |k,0\rangle 
\nonumber \\ 
& = & (1-|\zeta|^{2})^{k} \exp(\zeta K_{+}) |k,0\rangle \nonumber \\
& = & (1-|\zeta|^{2})^{k} \sum_{n=0}^{\infty} \left[ \frac{
\Gamma(n+2k) }{ n!\Gamma(2k) } \right]^{1/2} \zeta^{n} |k,n\rangle .
\label{22}     \end{eqnarray}
Here $\zeta = (\xi/|\xi|)\tanh|\xi|$, so $|\zeta|<1$. In the case of
the two-mode boson realization the SU(1,1) CS can be recognized as
well-known two-mode squeezed states with $\xi$ being a squeezing 
parameter \cite{WE}. Any intelligent state can be represented in the 
coherent-state basis \cite{BBA}. By using this analytic 
representation, we can find that a $K_{2}$-$K_{3}$ intelligent state 
is also coherent when its eigenvalue $\lambda$ is \cite{BBA}
\begin{equation}
\lambda = \pm ik \sqrt{\gamma^{2}+1} .
\label{23}      \end{equation}
The corresponding coherent-state amplitude $\zeta$ is real:
\begin{equation}
\zeta = \frac{1}{\gamma \pm \sqrt{\gamma^{2}+1}} .
\label{24}      \end{equation}
The condition $|\zeta|<1$ is satisfied if $\gamma > 0$ for upper sign
and $\gamma < 0$ for lower sign. Squeezing in $K_{3}$ ($|\gamma|>1$)
corresponds to values $|\zeta|<0.414$.

By using the definition (\ref{22}) of the SU(1,1) CS, one can easily
calculate the variance of $K_{2}$ \cite{WE}
\begin{equation}
(\Delta K_{2})^{2} = \frac{ k(1+|\zeta|^{4} - \zeta^{2}
- \zeta^{\ast 2}) }{ 2(1-|\zeta|^{2})^{2} }  .
\label{25}      \end{equation}
For a coherent-intelligent state, $\zeta$ is real, so we obtain
$(\Delta K_{2})^{2} = k/2$. Other expectation values over 
$|k,\zeta\rangle$ are 
\begin{equation}
(\Delta K_{3})^{2} = \frac{ 2k|\zeta|^{2} }{ (1-|\zeta|^{2})^{2} } ,
\label{26}      \end{equation}
\begin{equation}
\langle K_{1} \rangle = \frac{2k {\rm Re}\, \zeta}{1-|\zeta|^{2}} .
\label{27}      \end{equation}
Then it is straightforward to check that an equality is achieved in
the uncertainty relation (\ref{17}) provided that expectation values
are calculated over an SU(1,1) coherent state with real $\zeta$. 
It is seen that the states that belong to the coherent-intelligent 
intersection lead to the best measurement accuracy among all the 
SU(1,1) CS. The mean-square fluctuation in $\phi$ given by Eq.\ 
(\ref{19}) is, for the interferometer fed with an SU(1,1) 
coherent-intelligent state, 
\begin{equation}
(\Delta\phi)^{2} = \frac{1}{2k\sinh^{2}\!\beta} .
\label{28}      \end{equation}
We see that the phase sensitivity is independent of the value of
squeezing represented by $\zeta$. It depends only on parameter 
$\beta$ of the four-wave mixer and on the photon-number difference 
between the two input modes ($n_{0} = 2k-1$). Therefore, $\zeta$ can 
be taken to be zero, i.e., one can choose an input state with a fixed
number of photons in the one mode and the vacuum in the other. The 
value of $\beta$ is restricted by properties of available four-wave 
mixers. We see from Eq.\ (\ref{28}) that for a given value of 
$\beta$ the phase sensitivity of the SU(1,1) interferometers can be 
essentially improved by choosing input states with large values of 
the photon-number difference between the two modes. When $n_{0} = 0$ 
(in particular, when the vacuum enters both input ports), the phase 
fluctuations come to the known value 
$(\Delta\phi)^{2} = 1/\sinh^{2}\!\beta$.

It is usual to examine the interferometer efficiency by expressing
the phase sensitivity $\Delta\phi$ in terms of the total 
number $N$ of photons passing through the phase shifters. In the case
of the interferometer considered here, $N$ is the total number of
photons emitted by FWM1:
\begin{equation}
N = 2 \langle K'_{3} \rangle - 1 ,
\label{29}      \end{equation}
where ${\bf K}'=L(-\beta){\bf K}$, so $K'_{3} = (\cosh\beta) K_{3}
- (\sinh\beta) K_{2}$. By calculating the expectation value over a 
coherent-intelligent state, we obtain
\begin{equation}
N = 2k \frac{1+\zeta^{2}}{1-\zeta^{2}}\cosh\beta - 1 .
\label{30}      \end{equation}
Solving this equation for $\sinh^{2}\!\beta$ we finally find
\begin{equation}
(\Delta\phi)^{2} = \frac{1}{2k} \left[ \left( \frac{1-\zeta^{2}
}{1+\zeta^{2}}\, \frac{N+1}{2k} \right)^{2} - 1 \right]^{-1} .
\label{31}      \end{equation}
We see that the phase sensitivity $\Delta\phi$ approaches $1/N$. 
The best interferometer efficiency is achieved for $\zeta = 0$ and 
$n_{0} = 0$ ($k=1/2$). Then one gets the result for the vacuum 
input\footnote{Please note a minor difference between this result and
equation (9.31) of Ref. \cite{YMK} where `$-2$' is erroneously
printed instead of `$+2$'.} \cite{YMK}:
\begin{equation}
(\Delta\phi)^{2} = \frac{1}{N(N+2)} .
\label{32}      \end{equation}
For a given $N$, $(\Delta\phi)^{2}$ is optimized by taking the vacuum
in both input modes. However, for a given value of $\beta$
(dictated by practical considerations), the
phase sensitivity is improved by choosing input modes with
a large photon-number difference between them.

\begin{flushleft}

\end{flushleft}

\section*{Figure captions}

\noindent
Figure 1: An SU(1,1) interferometer. Two light modes $a_{1}$ and
$a_{2}$ are mixed by four-wave mixer FWM1, accumulate phase shifts
$\phi_{1}$ and $\phi_{2}$, respectively, and then they are again
mixed by four-wave mixer FWM2. The photons in output modes are
counted by detectors D1 and D2.

\end{document}